\documentclass[12pt]{article}
\usepackage{amsmath}

\newcommand{\eq}{\begin{equation}}
\newcommand{\eqx}{\end{equation}}
\newcommand{\eqn}{\begin{eqnarray}}
\newcommand{\eqnx}{\end{eqnarray}}
\newcommand{\alg}{\begin{align}}
\newcommand{\algx}{\end{align}}

\newcommand{\f}[2]{\frac{#1}{#2}}

\newcommand{\Lm}{\Lambda}

\newcommand{\al}{\alpha}
\newcommand{\bt}{\beta}
\newcommand{\om}{\omega}
\newcommand{\pl}{p_L}
\newcommand{\pt}{p_T}
\newcommand{\xl}{x_L}
\newcommand{\xt}{x_T}
\newcommand{\kl}{k_L}
\newcommand{\kt}{k_T}
\newcommand{\diag}[4]{
\left(\begin{array}{cccc}
#1 & 0 & 0&0 \\
0 & #2 & 0 & 0 \\
0 & 0 & #3 & 0 \\
0 & 0 & 0 & #4
\end{array}
\right)
}

\newcommand{\eps}{\varepsilon}
\newcommand{\qqqq}{\quad\quad\quad\quad}

\DeclareMathOperator{\arctanh}{arctanh}

\newcommand{\OO}[1]{{\cal O}\left(#1\right)}
\newcommand{\cor}[1]{\left\langle#1\right\rangle}

\title{Towards the description of anisotropic plasma at strong coupling}

\author{Romuald A. Janik\thanks{e-mail: {\tt ufrjanik@if.uj.edu.pl}}
\  and Przemys{\l}aw Witaszczyk\thanks{e-mail: {\tt bofh@th.if.uj.edu.pl}} \\ \\
Institute of Physics\\
Jagellonian University,\\
ul. Reymonta 4, \\
30-059 Krak{\'o}w\\
Poland
}

\begin{document}

\maketitle

\begin{abstract}
We initiate a study of anisotropic plasma at strong coupling using the AdS/CFT correspondence. We construct an exact dual geometry which represents a static uniform but anisotropic system and find, that although it is singular, it allows for a notion of `incoming' boundary conditions. We study small fluctuations around this background and find that the dispersion relation depends crucially on the direction of the wave-vector relative to the shape of the anisotropy reminiscent of similar behaviour at weak coupling. We do not find explicit instabilities to the considered order but only a huge difference in the damping behaviour.
\end{abstract}

\section{Introduction}

One of the outstanding problems in
our understanding of heavy-ion collisions is the question why quark-gluon plasma (QGP) can be described very accurately by hydrodynamics so soon after the
collision \cite{hydro}. The difficulty in addressing this question lies partly in a mixed weak/strong coupling physics of the initial state and early dynamics. Although some notable work has been done (see e.g. \cite{bottomup}), we are still far from a definite understanding.

Let us recall that hydrodynamics by definition involves the
notions of (isotropic) pressure while just after the collision the
energy-momentum tensor is definitely anisotropic.
In a first
approximation the question is thus to understand the process of
isotropisation of an anisotropic plasma system.

One of the mechanisms which was suggested to be responsible for this
behavior is the appearance of instabilities in an anisotropic plasma
system first discovered in \cite{Mrowczynski}. 
Since the study of an expanding (anisotropic) plasma system is quite
complicated if not impossible, a simpler system has been investigated
namely an anisotropic plasma sytem which fills the whole space and
evolves in Minkowski time (and not proper time).

The instablities at weak coupling have been investigated in detail \cite{Yaffe,RomStrik}. Subsequently the process of isotropisation has been studied in real time through numerical simulations \cite{Yaffe,Strik,Bodeker,BERGES}. Initially, for {\em weak fields} the evolution is exponential, in accordance with the instabilities discovered in weak coupling computations around an anisotropic plasma background, then when nonlinear field effects become stronger, the evolution becomes linear in time.

The motivation of this paper is to address similar issues at strong coupling using the AdS/CFT correspondence \cite{adscft} as a calculational tool. Our ultimate goal is to study the temporal evolution of the anisotropic plasma system and its approach to isotropy. However in the present paper we want to first investigate the situation when the anisotropic plasma is assumed to be static and to look for possible instabilities of small fluctuations in direct analogy to the weak coupling considerations. We plan to study the real-time isotropisation process in future work \cite{FUTURE}.

The plan of this paper is as follows. In section 2, we will briefly review some features of plasma instabilites at weak coupling. In section 3 we review the AdS/CFT framework used and in section 4 we construct the geometry dual to a static anisotropic plasma system. In the following two sections we comment on some of its pathologies and discuss the issue of defining physically natural boundary conditions. In section 7 we study the dispersion relation of R-charge fluctuation modes and we close the paper with a discussion and an appendix containing the relevant wave equations.

\section{Plasma instabilities at weak coupling}

In this section we will briefly describe some qualitative features of plasma instabilities studied at weak coupling. The majority of numerical and analytical work has been done for an infinite, spatially uniform system which does not expand. 
One basically starts from some initial conditions when momentum distributions are anisotropic.
Typically one has a
separation between hard and soft modes and studies how the hard modes
lead to isotropisation of the soft modes but this is not strictly
necessary \cite{BERGES}.
Then one studies the time dependence of electric/magnetic fields
and consequently the time evolution of the energy-momentum tensor:
\eq
\label{e.timedep}
T_{\mu\nu}=\diag{\eps}{\pl(t)}{\pt(t)}{\pt(t)}
\eqx
where $\eps=\pl+2\pt$. Such studies have to involve numerical
computations.

Alternatively, one can try to compute the poles of the gluon
propagator in the anisotropic system with a momentum distribution
\eq
f(\vec{p})=\sqrt{1+\xi}\, f_{iso}(\vec{p}^2+\xi \pl^2)
\eqx
where $f_{iso}$ is some isotropic distribution and $\xi$ is the
anisotropy parameter related to the ratio of transverse to longitudinal
pressure through
\eq
\label{e.xi}
\xi=\f{\pt}{\pl}-1
\eqx
The outcome is that some modes develop unstable
behavior. The precise behavior crucially depends on the {\em sign}
of the anisotropy. If $\xi>0$, then all modes with
transverse wave vectors remain stable, while the
longitudinal ones develop an instability in a finite range of $\kl$. If $\xi<0$
the situation is reversed with the longitudinal modes remaining stable and the
transverse modes developing an instability.

The unstable modes can be roughly identified with the initial
behavior of the plasma in the time-dependent simulations outlined above. Thus the
simpler computation of the modes gives an indication of the direction
of the evolution of the system. Later when nonlinear
effects become important, the evolution ceases to be exponential.

In the strong
coupling regime one may attempt to study these questions in both
ways. In this paper we will analyze what happens when we try to mimick
the simpler approach, namely to consider an anisotropic system and
look at small fluctuations. We plan to analyze the more realistic case
of time dependence in the future.

\section{The AdS/CFT framework}

Within the AdS/CFT correspondence, a system of plasma is described by a
dual geometry which is constructed as follows. Let us assume that the
only nonvanishing expectation value in the system is a certain
profile of the energy-momentum tensor:
\eq
\cor{T_{\mu\nu}(x^\mu)}
\eqx
which has a specific dependence on the Minkowski coordinates
$x^\mu$. Then the dual geometry is written in the Fefferman-Graham
coordinates as
\eq
ds^2=\f{g_{\mu\nu}(x^\mu,z) dx^\mu dx^\nu+dz^2}{z^2}
\eqx
The metric is determined by solving Einstein's equations with negative
cosmological constant ($\Lm=-6$) which can be written as
\eq
\label{e.einst}
R_{\al\bt}+4g_{\al\bt}=0
\eqx
with the boundary condition at $z=0$ given by
\eq
g_{\mu\nu}(x^\mu,z)=\eta_{\mu\nu}+z^4 g^{(4)}_{\mu\nu}(x^\mu)+\OO{z^6}
\eqx
where $g^{(4)}_{\mu\nu}(x^\mu)$ is related to the expectation value of
the gauge theoretical energy-momentum tensor \cite{Skenderis}:
\eq
\cor{T_{\mu\nu}(x^\mu)} =\f{N_c^2}{2\pi} g^{(4)}_{\mu\nu}(x^\mu)
\eqx

The proposal put forward in \cite{US} is to determine the geometry for
a given energy-momentum profile and choose the physical one by
requiring the nonsingularity of the resulting bulk geometry.

This framework is geared to study the time-dependent processes and
thus can be used to study isotropisation in real time starting from
the time-dependent energy-momentum tensor (\ref{e.timedep})
by constructing the dual geometry and requiring nonsingularity.
We plan to follow this route in future work \cite{FUTURE}.

Because the above procedure, although conceptually simple, is
technically quite involved, we decided to analyze what are the
fluctuations around a system where we neglect the time dependence of
the anisotropy of the plasma. This problem is a direct analog of the
computation of the (unstable) poles of the gluon propagator in the
unstable medium. Similarly as in the weak coupling case we do not
expect such a static system to exist indefinitely. Certainly even if
the system would be classically fine-tuned to stay in the unstable
regime, quantum fluctuations would cause it to evolve. Therefore we
expect the geometry to be somewhat pathological. The question that we
wanted to ask is whether any kind of information may be extracted from
it and see whether the pattern of fluctuation modes has some
resemblance to the weak coupling situation.

\section{Geometry dual to a static anisotropic plasma system}

Let us now find the dual geometry corresponding to a uniform, static
and anisotropic energy-momentum tensor:
\eq
\cor{T_{\mu\nu}}=\diag{\eps}{\pl}{\pt}{\pt}
\eqx
with $\eps=\pl+2\pt$.
The most general metric with these symmetries has the form
\eq
\label{e.geom}
ds^2=\f{1}{z^2} \left( -a(z) dt^2 +b(z) d\xl^2 +c(z)d\xt^2 +dz^2 \right)
\eqx
The scalar functions $a(z)$, $b(z)$ and $c(z)$ have to vanish at
$z=0$, and their $z^4$ coefficients are related to the transverse and
longitudinal pressure. The general solution of (\ref{e.einst}) satisfying these constraints reads
\eqn
a(z) &=& (1+A^2z^4)^{\f{1}{2}-\f{1}{4}\sqrt{36-2B^2}}
         (1-A^2z^4)^{\f{1}{2}+\f{1}{4}\sqrt{36-2B^2}}\\
b(z) &=&  (1+A^2z^4)^{\f{1}{2}-\f{B}{3}+\f{1}{12}\sqrt{36-2B^2}}
         (1-A^2z^4)^{\f{1}{2}+\f{B}{3}-\f{1}{12}\sqrt{36-2B^2}}\\
c(z) &=&  (1+A^2z^4)^{\f{1}{2}+\f{B}{6}+\f{1}{12}\sqrt{36-2B^2}}
         (1-A^2z^4)^{\f{1}{2}-\f{B}{6}-\f{1}{12}\sqrt{36-2B^2}}
\eqnx
and the $A$ and $B$ parameters are related to the energy density and
pressure through
\eqn
\eps &=&\f{1}{2} A^2 \sqrt{36-B^2} \\
\pl &=&\f{1}{6} A^2 \sqrt{36-B^2} -\f{2}{3} A^2 B \\
\pt &=&\f{1}{6} A^2 \sqrt{36-B^2} +\f{1}{3} A^2 B
\eqnx
It is also convenient to link the $B$ parameter with the $\xi$
anisotropy parameter defined through (\ref{e.xi}) to get
\eq
B=\f{6 \xi}{\sqrt{18 \xi^2+48 \xi+36}}
\eqx
When there is no anisotropy, $B=0$ and the above solution reduces to
the standard static AdS black hole solution.

\section{The issue of the singularity}

An unavoidable property of the metric is, that once we have nonzero
anisotropy, there is a singularity in the bulk, which is a very
significant obstacle towards making a physical interpretation of our
setup. This can be interpreted that a static anisotropic plasma system
cannot exist at strong coupling. In fact such a statement is also
true at weak coupling as discussed in section~2.

We would like to try to interpret our setup as a snapshot of the
dynamical evolving scenario (close to the initial condition) and to
investigate small fluctuations around such a system in order,
eventually, to compare with a similar setup at weak coupling where the
pattern of stable/unstable modes has a specific dependence on the sign
of anisotropy and the relative orientation of the wave-vectors.

This might give a hint towards the real time-dependent evolution of
the anisotropic system and what would be the differences in behavior
with what is known at weak coupling before attempting a real dynamical
computation at strong coupling.

Let us note, however, that because of the appearance of the
singularity it is not clear what is the regime of validity of these
results. The `snapshot' interpretation could perhaps be validated by assuming
very big `classical' occupation numbers but we do not know how to
estimate possible timescales appearing at strong coupling. We decided
to proceed nevertheless with the computation of small fluctuations and
to see {\em a-posteriori} whether the observed behavior is physically
viable, or whether it is obviously pathological.

\section{Boundary conditions}

One of the major conceptual issues that one has to face when
considering spacetimes with naked singularities is the problem of
what boundary conditions to impose at the singularity. This problem
has been considered in the general relativity literature \cite{Gibbons,negmass}, and not
surprisingly the problem of stability/instability of such spacetimes
like the negative mass Schwarzschild black holes depends on the chosen
boundary conditions.

In fact even for the ordinary Schwarzschild geometry if one would
choose outgoing boundary conditions at the horizon we would
get unstable modes. This issue of course never arises in the case
of the black hole horizon, as there incoming boundary conditions are clearly singled out
physically.

We will try to find an analog of the incoming boundary conditions for
our geometry and use it for our computations. In fact the geometry is
not so pathological as e.g. a negative mass black hole where a similar
construction would not be possible. This feature of (\ref{e.geom}) is
somewhat reassuring.

Let us consider the wave equation for a massless scalar field in the
geometry (\ref{e.geom}). For simplicity we will set $A=1$.
Performing the standard separation of
variables
\eq
\Phi=\phi(z) e^{-i \om t+i k_1 x^1 +i k_3 x^3}
\eqx
and a subsequent change of variables
\eq
x=\f{1}{4} \arctanh z^4
\eqx
we obtain the scalar equation in the form
\eqn
&&\f{d^2\phi}{dx^2} + \f{8}{(e^{16x}-1)^{\f{3}{2}}} \biggl( \om^2
e^{2(6+\sqrt{36-2B^2})x} -\kl^2
e^{2(6+\f{4B}{3}-\f{1}{3}\sqrt{36-2B^2})x}+\nonumber\\
&&\qqqq\qqqq -\kt^2 e^{2(6-\f{2B}{3}-\f{1}{3}\sqrt{36-2B^2})x} \biggr)\phi=0
\eqnx
where the singularity is at $x=\infty$. We see that close to the
singularity the piece proportional to $\om^2$ dominates and the
equation looks there as
\eq
\label{e.asym}
\f{d^2\phi}{dx^2} + 8\om^2 e^{-2(6-\sqrt{36-2B^2})x}  \phi=0
\eqx
For zero anisotropy $B=0$ this has the familiar $e^{- i \sqrt{8} \om x}$,
$e^{+i \sqrt{8} \om x}$ solutions which correspond to incoming and outgoing
waves respectively. If we would consider the limit of very small $B$,
we see that the solution behaves as for an ordinary horizon, with a
clear separation of incoming and outgoing waves, which would only be
modified very close to the singularity. In fact linearizing the
equation in $B$ will reduce the boundary conditions to the black hole ones.
For general $B$, the solution of (\ref{e.asym}) is a combination of
Hankel functions
\eq
H_0^{1,2}\left( \f{\sqrt{8}}{C} \om e^{-C x} \right)
\eqx
with $C=6-\sqrt{36-2B^2}$.
As is well known Hankel functions form a convenient basis of incoming
and outgoing wavefunctions in a cylindrical geometry. Hence in general
we also have a clear definition of incoming and outgoing waves.

Since we do not want any
information flow from the singularity into the bulk, we will always pick
these `generalized' incoming waves as our boundary condition at the
singularity.

Let us note that the geometry (\ref{e.geom}) is rather special in that
it allows for such a choice to exist at all. If we were to perform the same
analysis, say for a negative mass (planar) AdS black hole we would
have found that for nonzero wave-vectors these terms would dominate the
$\om^2$ term and give solutions for which there would be no notion of an
incoming/outgoing wave interpretation.

\section{R-charge fluctuation modes}

In view of the appearance of plasma instabilities at weak coupling we
are interested mainly in studying low lying fluctuation modes of the
geometry (\ref{e.geom}). Let us therefore briefly review the situation
for isotropic plasma \cite{Son1,Son2,KS}. There all the modes exhibit damping (quasinormal
modes). The smallest damping is associated with modes in the
hydrodynamical regime. These involve modes of the graviton in the
shear and sound channels which have the dispersion relations
\eq
\om =-i \f{\eta}{\eps+p} k^2+\ldots \qqqq \om=\f{1}{\sqrt{3}}
k-i\f{2}{3}\eta \f{k^2}{\eps+p} +\ldots
\eqx
respectively,
as well as modes of the bulk gauge field associated with correlation
functions of R-charge currents
\eq
\label{e.rdisporg}
\om=-i D_R k^2+\ldots
\eqx
where $D_R=1/2\pi T$ is the R-charge diffusion constant. All other
modes have finite non-zero damping ($Im\, \om<const<0$) even for very long
wavelength modes \cite{Hubeny,Starinets,KS}.

If we are to see an instability we have therefore to concentrate on
the modes for which the (negative) imaginary part is smallest i.e. on
the hydrodynamic modes.
In the present paper we will study the bulk gauge fluctuations
as these are technically simplest. In the major part of the paper we
will set $A=1$ and reinstate it in the final results and formulas in
the appendices. In these units the dispersion relation for
R-charge diffusive modes (\ref{e.rdisporg}) for {\em isotropic plasma}
i.e. with $B=0$ takes the form
\eq
\label{e.dispiso}
\om=-i \f{k^2}{2\sqrt{2}}+\ldots
\eqx
In this section we will study how this dispersion relation is modified
by the presence of anisotropy.

The equation of motion for the bulk gauge field takes the form
\eq
\partial_\al \left( \sqrt{-g} F^{\al \bt} \right)=0
\eqx
Due to the anisotropy of the plasma configuration and hence of the
bulk geometry (\ref{e.geom}), one of the spatial coordinates is
singled out (the longitudinal one), we will denote it by $y$, while
the two transverse ones will be denoted by $x_{1,2}$. Similarly we
will have a decomposition for wave vectors in the Fourier
transform of the vector potential,
\eq
A_{\mu}(x)=\int\frac{d^{3}k d\om}{(2\pi)^{4}}e^{-i\om t+i\vec{k}\vec{x}}A_{\mu}(z,k)
\eqx
We will separately study the extreme situations when the wave vector
is purely longitudinal or purely transverse:
\eq
L:\ k=(\kl,0,0),\ T: q=(0,0,\kt)
\eqx
Let us recall that at weak coupling only one type of these modes
develops an instability depending on the sign of the anisotropy parameter.

Furthermore, since we are dealing with vector fields
so there will be modes of the fields parallel and orthogonal to the
wave vector.
As a result we get two sets of coupled differential equations which can be
further simplified by introducing gauge invariant variables, different
in each set:\\
Longitudinal modes:
\begin{align}\textrm{(L-L)}\ &E_{y}(\kl,z)=\omega A_{y}(\kl,z)+
  \kl A_{t}(\kl,z),\ \kl||E_{y}\\
\textrm{(L-T)}\ &E_{1}(\kl,z)=\omega
A_{1}(\kl,z),\ E_{2}(\kl,z)= \omega A_{2}(\kl,z)
\end{align}
Transverse modes:
\begin{align}\textrm{(T-T)}\ &E_{1}(\kt,z)=\omega A_{1}(\kt,z)+
  \kt A_{t}(k,z),\ \kt||E_{1},\\
\textrm{(T-L)}\ &E_{y}(\kt,z)=\omega
A_{y}(\kt,z),\ E_{2}(\kt,z)= \omega A_{2}(\kt,z)
\end{align}
After that we are left with five quite lengthy equations. In the
isotropic limit of the standard static black hole it is the electric modes
which are parallel to the wave-vector which exhibit diffusive
behavior (here (T-T) and (L-L)). The rest have $\OO{1}$ damping even
for small $k$.

In the present paper we will be mostly interested in effects appearing
for small anisotropy, hence we will linearize the resulting equations
in $B$. We quote all resulting equations in this regime in Appendix A.

Here we will analyze the equations for (T-T) and (L-L) setting $A=1$,
reinstating $A$ dependence in the final answer.

Close to the singularity $z=1$, in the approximation linearized in $B$ we
find the following asymptotic behaviors which can be associated with
incoming boundary conditions
\eq
E_y \sim (1-z)^{-i \f{\om}{2\sqrt{2}} +\f{B}{6}} \qqqq
E_1 \sim (1-z)^{-i \f{\om}{2\sqrt{2}} -\f{B}{12}}
\eqx
Let us now concentrate on the equation for $E_y$ and make a
decomposition
\eq
E_y(z)=(1-z)^{-i \f{\om}{2\sqrt{2}} +\f{B}{6}}g(u)
\eqx
where we used the variable $u=z^2$. Furthermore, as we are interested
in small frequencies and wave vectors we will rescale $\om \to \eps
\om$ and $\kl \to \eps \kl$ and write
\eq
g(u)=1+\eps g_0^a(u)+\eps^2 g_0^b(u) +B (g_1^a(u)+\eps g_1^b(u)+\ldots)+\ldots
\eqx
We then impose vanishing conditions at $u=1$ for the $g_i^A(u)$'s
which determines them completely. At this stage the mode has the
correct incoming boundary condition at the horizon. Imposing Dirichlet
conditions at the boundary gives the dispersion relation for the
modes:
\eq
\label{e.dirichlet}
g(0)=1+g_0^a(0)+g_0^b(0)+B(g_1^a(0)+g_1^b(0))=0
\eqx
We find for the leading terms
\eqn
\label{e.0a}
g_0^a(0)&=& \f{i \kl^2}{2\sqrt{2} \om} -\f{i \log 2}{2\sqrt{2}} \om \\
\label{e.1a}
g_1^a(0) &=& -\f{\kl^2}{6\om^2} +\f{\log 4}{12}
\eqnx
and for the subleading ones
\eqn
\label{e.0b}
g_0^b(0) &=& -\left( \f{\pi^2}{96}+\f{\log^2 2}{16} \right) \om^2
+\f{\log 2}{8} \kl^2 \\
\label{e.1b}
g_1^b(0) &=& i \f{\sqrt{2}}{288} (\pi^2-12\log^2 2) \om
\eqnx
Let us analyze the condition (\ref{e.dirichlet}). For zero $B$, we
recover immediately (\ref{e.dispiso}) from (\ref{e.0a}). Once we turn
on a small anisotropy, we find that for small $\kl$ we cannot sustain
the scaling $\om \propto k^2$. Indeed then the first term in
(\ref{e.1a}) will start to dominate and we get essentially
\eq
1 -\f{\kl^2}{6\om^2} B=0
\eqx
Here we see that the outcome depends crucially on the {\em sign} of
the anisotropy. If $B>0$ then, although we get no instability, the
damping vanishes to this order and one has
\eq
\om=\sqrt{\f{B}{6}} \kl+\ldots
\eqx
On the other hand if $B<0$ we get very strong damping\footnote{The
  sign is chosen as to reduce to the isotropic case when taking more
  terms into account and performing $B\to 0$.}
\eq
\label{e.negb}
\om=-i\sqrt{\f{-B}{6}} \kl+\ldots
\eqx
When we increase $\kl$ the first term in (\ref{e.0a})-(\ref{e.1b})
that will start to become important will be the first term in
(\ref{e.0a}) when $\kl$ becomes of of order $\sqrt{|B|}$. At this
stage we have to take into account both terms and we get
\eq
1 -\f{\kl^2}{6\om^2} B+ \f{i \kl^2}{2\sqrt{2} \om}=0
\eqx
This is a quadratic equation which can be solved to give (after inserting dependence on $A$ which was previously set to 1) 
\eq
\om= -i \f{\kl^2+ \sqrt{\kl^4-\f{16}{3}A^{\frac{1}{4}}B \kl^2}}{4\sqrt{2}A^{\frac{1}{8}}}
\eqx
We see that for $B<0$, the sign of the square root has to be chosen as
above in order to reproduce smoothly the $B\to 0$ limit. This
justifies the choice of sign in (\ref{e.negb}). For positive $B$ of
course both signs are possible and there is a branch point on the real
axis.

Let us note that the change in the qualitative behavior depending on
the sign of the anisotropy is quite similar to the one seen at weak
coupling. For $B>0$, at weak coupling modes with longitudinal wave
vectors develop an instability while for $B<0$ they remain stable.
Here the longitudinal ones develop a linear regime for $B>0$ while for $B<0$ they
exhibit very strong damping.

For modes with nonzero transverse modes the situation is reversed. The equation that
gives the dispersion relation is:

\eq
1 +\f{\kt^2}{12\om^2} B+ \f{i \kt^2}{2\sqrt{2} \om}=0,
\eqx
and the solution reads (after reinstating the dependence on $A$):
\eq
\om= -i \f{\kt^2+ \sqrt{\kt^4+\f{8}{3}BA^{\frac{1}{4}} \kt^2}}{4\sqrt{2}A^{\frac{1}{8}}}.
\eqx
In this case the term with $B$ differs by a numeric constant but, what is more important, also has a different sign. This means that the behaviour of the modes with transverse wave vectors has an opposite pattern of strong/weak damping relative to the {\em sign} of the anisotropy. This is qualitatively similar to the weak coupling behaviour reviewed in section 2.

One can also compute two-point functions using the standard procedure and find that the poles coincide with the dispersion relations found above.

\section{Discussion}

In the present paper we have studied the properties of a uniform infinite plasma system with anisotropic pressure. We have constructed the corresponding dual geometry and found that it is singular. This shows that such a system cannot be considered to exist indefinitely. The singularity is relatively mild in the sense that for a scalar wave equation in this background geometry there is a natural notion of ingoing boundary conditions in contrast to generic singular geometries. This feature allowed us to investigate small fluctuations and to look for possible instabilities and whether they would manifest themselves at the linear level in the supergravity description.

Similarly as for weak coupling, we find that the qualitative behaviour of longitudinal and transverse modes depends crucially on the sign of anisotropy.
In contrast to weak coupling however, to leading order in the anisotropy parameter, we do not find instabilities but massless-like propagation (albeit with some quadratic damping) for modes with wave vectors for which one would expect instabilities at weak coupling, and very strong damping for modes which would remain stable at weak coupling. This qualitative correlation between the sign of the anisotropy and behaviour of the modes is reassuring in view of our doubts on using the singular geometry at all. Perhaps the lack of instabilities seen at the linearized level might be analogous to the lack of exponential growth for strong fields in the numerical simulations at weak coupling. In addition, Boltzmann-Vlasov simulations with large collision rate (corresponding to strong coupling) also showed the disappearance of instabilities \cite{MS}.

One could extend the present investigation by going to higher orders in the anisotropy parameter but the most interesting question in our opinion is to determine the real-time evolution of the anisotropic system at strong coupling. This dynamical problem, although technically much harder, would not be plagued by conceptual issues dealing with singularities. It would be very interesting to confront the outcome with the linearized results of the present paper. We plan to investigate the real-time evolution in future work~\cite{FUTURE}.

\bigskip

\noindent{}{\bf Acknowledgements.} We would like to thank P. Bizon and A. Rostworowski for discussions on quasi-normal modes. RJ would like to thank J. Berges, S. Mrowczynski, M. Strickland, L. Yaffe for illuminating discussions during the KITP program `Nonequilibrium Dynamics in Particle Physics and Cosmology' and acknowledge support by the National Science Foundation under Grant No. PHY05-51164. 
This work was supported in part by the Polish Ministry of Science and
Information Technologies grant 1P03B04029 (2005-2008), RTN network
ENRAGE MRTN-CT-2004-005616, and the Marie Curie ToK COCOS (contract MTKD-CT-2004-517186).

\appendix

\section{Linearized equations for R-charge modes}

In the following equations we neglected {\em quadratic} terms in $B$.

\medskip

\noindent{\bf Longitudinal wave-vectors}
\begin{align}
&E_{1}''-\frac{3+4\sqrt{A}(3+B)z^{4}+9Az^{8}}{3z(1-Az^{8})}E_{1}'\\
&+(\frac{\om^{2}(1+\sqrt{A}z^{4})}{(1-\sqrt{A}z^{4})^{2}}-k_{L}^{2}(1-\sqrt{A}z^{4})^{-\frac{B}{3}}(1+\sqrt{A}z^{4})^{-1+\frac{B}{3}})E_{1}=0,
\end{align}
\begin{align}
&E_{y}''-\frac{3k_{L}^{2}(1-\sqrt{A}z^{4})^{2}(1+\sqrt{A}z^{4})^{\frac{B}{3}}(1-3\sqrt{A}z^{4}(4-\sqrt{A}z^{4}))}
{3z(1-Az^{8})(k_{L}^{2}(1-\sqrt{A}z^{4})^{2}(1+\sqrt{A}z^{4})^{\frac{B}{3}}-\om^{2}(1-\sqrt{A}z^{4})^{\frac{B}{3}}(1+\sqrt{A}z^{4})^{2})}+\\
&\frac{\om^{2}(1-\sqrt{A}z^{4})^{\frac{B}{3}}(1+\sqrt{A}z^{4})^{2}(3+\sqrt{A}z^{4}(12-8B+9\sqrt{A}z^{4}))}
{3z(1-Az^{8})(k_{L}^{2}(1-\sqrt{A}z^{4})^{2}(1+\sqrt{A}z^{4})^{\frac{B}{3}}-\om^{2}(1-\sqrt{A}z^{4})^{\frac{B}{3}}(1+\sqrt{A}z^{4})^{2})}\\
&+(\frac{\om^{2}(1+\sqrt{A}z^{4})}{(1-\sqrt{A}z^{4})^{2}}-k_{L}^{2}(1-\sqrt{A}z^{4})^{-\frac{B}{3}}(1+\sqrt{A}z^{4})^{-1+\frac{B}{3}})E_{y}=0
\end{align}

(equation for $E_{2}$ is exactly the same as for the $E_{2}$).

\medskip

\noindent{\bf Transverse wave-vectors}
\begin{align}
&E_{2}''-\frac{3+4\sqrt{A}(3+B)z^{4}+9Az^{8}}{3z(1-Az^{8})}E_{2}'\\
&+(\frac{\om^{2}(1+\sqrt{A}z^{4})}{(1-\sqrt{A}z^{4})^{2}}-k_{T}^{2}(1-\sqrt{A}z^{4})^{-\frac{B}{6}}(1+\sqrt{A}z^{4})^{-1-\frac{B}{6}})E_{2}=0,
\end{align}

\begin{align}
&E_{y}''-\frac{3+\sqrt{A}z^{4}(12-8B+9\sqrt{A}z^{4})}{3z(1-Az^{8})}E_{y}'\\
&+(\frac{\om^{2}(1+\sqrt{A}z^{4})}{(1-\sqrt{A}z^{4})^{2}}-k_{T}^{2}(1-\sqrt{A}z^{4})^{-\frac{B}{6}}(1+\sqrt{A}z^{4})^{-1-\frac{B}{6}})E_{y}=0,
\end{align}

\begin{align}
&E_{1}''+(\frac{-3k_{T}^{2}(1-\sqrt{A}z^{4})^{2+\frac{B}{6}}(1-3\sqrt{A}z^{4}(4-\sqrt{A}z^{4}))}{3z(1-Az^{8})(k_{T}^{2}(1-\sqrt{A}z^{4})^{2+\frac{B}{6}}
-\om^{2}(1-\sqrt{A}z^{4})^{2+\frac{B}{6}})}+\\
&\frac{\om^{2}(1+\sqrt{A}z^{4})^{2+\frac{B}{6}}(3+4\sqrt{A}(3+B)z^{4}+9Az^{8})}{3z(1-Az^{8})(k_{T}^{2}(1-\sqrt{A}z^{4})^{2+\frac{B}{6}}
-\om^{2}(1-\sqrt{A}z^{4})^{2+\frac{B}{6}})})E_{1}'\\
&+(\frac{\om^{2}(1+\sqrt{A}z^{4})}{(1-\sqrt{A}z^{4})^{2}}-k_{T}^{2}(1-\sqrt{A}z^{4})^{-\frac{B}{6}}(1+\sqrt{A}z^{4})^{-1-\frac{B}{6}})E_{1}=0
\end{align}

\end{document}